\documentclass[twocolumn,superscriptaddress,preprintnumbers,amsmath,amssymb,pra]{revtex4-2}

\usepackage{graphicx}
\usepackage{dcolumn}
\usepackage{bm}
\usepackage{ifpdf}
\usepackage{epstopdf}
\usepackage{slashed}
\usepackage{amsfonts}
\usepackage{mathrsfs}
\usepackage{amssymb}
\usepackage{multirow}
\usepackage{CJK}
\usepackage{xcolor}
\usepackage{textcomp}
\usepackage{csquotes}
\usepackage{tikz}
\usepackage{tikz-feynman}
\tikzfeynmanset{compat=1.1.0}

\def\lesssim{\ \raise.3ex\hbox{$<$}\kern-0.8em\lower.7ex\hbox{$\sim$}\ }
\def\gesim{\ \raise.3ex\hbox{$>$}\kern-0.8em\lower.7ex\hbox{$\sim$}\ }

\begin{document}
\begin{CJK}{UTF8}{ipxm}

\title{Comparative study of quartet superfluid state: \\ Quartet Bardeen-Cooper-Schrieffer theory and generalized Nambu-Gor'kov formalism}

\author{Yixin Guo (郭一昕)}
\email{yixin.guo@riken.jp}
\affiliation{RIKEN Nishina Center for Accelerator-Based Science, Wako 351-0198, Japan}

\author{Hiroyuki Tajima (田島裕之)}
\email{htajima@g.ecc.u-tokyo.ac.jp}
\affiliation{Department of Physics, Graduate School of Science, The University of Tokyo, Tokyo 113-0033, Japan}
\affiliation{Quark Nuclear Science Institute, The University of Tokyo, Tokyo 113-0033, Japan}

\author{Haozhao Liang (\CJKfamily{min}{梁豪兆})}
  \email{haozhao.liang@phys.s.u-tokyo.ac.jp}
  \affiliation{Department of Physics, Graduate School of Science, The University of Tokyo, Tokyo 113-0033, Japan}
  \affiliation{Quark Nuclear Science Institute, The University of Tokyo, Tokyo 113-0033, Japan}
  \affiliation{
    RIKEN Center for Interdisciplinary Theoretical and Mathematical Sciences (iTHEMS),
    Wako 351-0198, Japan}

\date{\today}

\begin{abstract}
We theoretically investigate a quartet superfluid state in fermionic matter by using the quartet Bardeen-Cooper-Schrieffer (BCS) variational theory and the Green's function method. 
We demonstrate that the quartet BCS theory with the multiple-infinite-product ansatz successfully reproduces an exact four-body result in a one-dimensional four-component Fermi gas at the dilute limit, in contrast to the single-infinite-product ansatz. 
To see the validity of the quartet BCS state, we derive the self-consistent equation for the quartet superfluid order parameter within the generalized imaginary-time Nambu-Gor'kov formalism, which is found to be consistent with the quartet BCS variational equation.
Moreover, by numerically computing the momentum-resolved single-particle spectral function in a one-dimensional system, we discuss how the single-particle spectra evolve with increasing the strength of the four-body cluster formation.
We show that a coherent BCS-like quasiparticle branch on the weak-coupling side evolves into a strongly damped, continuum-dominated spectrum in the strong-coupling side, while nonzero quartet superfluid order parameter persists throughout the crossover regime.
Our results would be useful for understanding beyond-BCS pairing effects and four-body cluster formations in fermionic systems in an interdisciplinary way.
\end{abstract}

\maketitle

\section{Introduction}\label{sec:I}

In nuclear many-body systems, the emergence of clustering phenomena plays a crucial role in understanding the structure and dynamics. 
While the conventional Bardeen-Cooper-Schrieffer (BCS) theory~\cite{Bardeen1957Phys.Rev.108.1175--1204} successfully describes the formation of different kinds of Cooper pairs and superfluidity in nuclear matter~\cite{10.1143PTP.112.27, 
RevModPhys.75.607} and finite nuclei~\cite{ring2004nuclear}, their application range remains limited to two-body correlations. 
However, in low-density regimes, particularly near the nuclear surface of heavy nuclei, four-body $\alpha$-particle-like quartets are expected to dominate the pairing mechanism~\cite{Tanaka2021Science.371.260}. 
These quartet correlations represent an essential extension of the pairing concept, capturing the strong binding and collective behavior characteristic of $\alpha$ clustering.

Consequently, many-body theories involving cluster formations beyond the BCS paradigm, as well as the liquid-drop model related to four-body quartet correlations, have been studied~\cite{Tohsaki2001Phys.Rev.Lett.87.192501,JSuper2010,Sogo2010Phys.Rev.C81.064310,PhysRevC.103.024316,PhysRevLett.131.193401}. 
For instance, the variational ansatz called the quartet condensation model (QCM) based on the multiple occupation of quartet operators has been proposed~\cite{FLOWERS1963586,Sandulescu2012Phys.Rev.C85.061303}.
The quartet condensation has been generalized to the coherent state ansatz in analogy with the BCS theory, which is called quartet BCS theory~\cite{Sen'kov2011Phys.Atom.Nuclei74.1267,Baran2020Phys.Lett.B805.135462,Baran2020Phys.Rev.C102.061301,Guo2022Phys.Rev.C105.024317,Guo2022Phys.Rev.Research4.023152,Guo2025Phys.Rev.C112.024310}. 
In Refs.~\cite{Sen'kov2011Phys.Atom.Nuclei74.1267,Guo2022Phys.Rev.C105.024317,Guo2022Phys.Rev.Research4.023152}, it has been shown that the quartet BCS theory provides a systematic framework for generalizing the BCS formalism to include four-fermion correlations in the infinite nuclear matter.
The quartet BCS theory aims to unify the treatment of pairing and quarteting within a coherent theoretical structure, and it has been also applied to finite nuclei~\cite{Baran2020Phys.Lett.B805.135462,Baran2020Phys.Rev.C102.061301,Guo2025Phys.Rev.C112.024310}
and condensed-matter systems~\cite{kamei2005quartet,grinenko2021state,PhysRevLett.131.193401,PhysRevResearch.6.033171}.
While the quartet coherent state considered in the quartet BCS theory breaks the particle-number conservation even in finite nuclei, the results of the correlation energy agree well with those with those of QCM that conserves particle number~\cite{Baran2020Phys.Lett.B805.135462}.
On the other hand, in addition to the issue about the broken number conservation, the quartet BCS theory has an ambiguity associated with the form of the coherent states at each quantum number in contrast to the BCS pairing state~\cite{Sen'kov2011Phys.Atom.Nuclei74.1267,Guo2022Phys.Rev.C105.024317,Guo2022Phys.Rev.Research4.023152}.
In this regard, testing the validity of the ansatz used in the quartet BCS theory is of great importance, as it could bridge the gap between microscopic many-body theories and the phenomenological understandings of cluster formation in nuclear systems.

Apart from the  variational approach, the Nambu-Gor'kov formalism~\cite{Gorkov195934,Gorkov195936,Nambu1960Phys.Rev.117.648--663} offers a systematic field-theoretical framework for describing superconductivity and superfluidity. 
Originally introduced by Nambu and Gor'kov in the late 1950s, this formalism reformulates the many-body problem in terms of two-component (particle-hole) Green's functions, allowing both normal and anomalous propagators to be treated on an equal footing. 
The Nambu-Gor'kov formalism thus provides a natural way to incorporate spontaneous symmetry breaking and the emergence of collective excitations in fermionic systems~\cite{schrieffer2018theory}. 
Extending this formalism to include the quartet degrees of freedom allows one to generalize the conventional pairing picture and explore the coexistence and competition between pairing and quarteting correlations within a unified Green's function framework~\cite{Sogo2010Phys.Rev.C81.064310,schuck2012quartetting}.
On the other hand, the relationship between the quartet BCS theory and the Nambu-Gor'kov formalism has been elusive due to their complexities in the application to nuclear systems.

In addition to its relevance to $\alpha$-like clustering in nuclear systems, a one-dimensional four-component Fermi gas provides a unique and theoretically well-controlled platform for studying quartet correlations~\cite{SCHUCK2007285}.
It is known from the exact Bethe-ansatz solutions of attractive SU(N) Fermi gases that, in one dimension, fermions form tightly bound $N$-body clusters in the ground state, so that for SU(4) the elementary bound object is a four-fermion quartet rather than a Cooper pair~\cite{PhysRevLett.20.98,takahashi1999,RevModPhys.85.1633}. 
These quartets appear as four-body string solutions and dominate the low-energy physics in the attractive regime. 
From an effective-field-theory perspective, the low-energy physics of one-dimensional four-component Fermi gases at unitarity can be captured by a universal four-body interaction, providing a controlled framework for studying quartet correlations beyond two-body physics~\cite{Nishida2010Phys.Rev.A82.043606,PhysRevLett.109.250403,PhysRevA.85.063624,PhysRevA.87.063617}.
Such multicomponent one-dimensional systems can be realized experimentally using ultracold atomic gases with multiple hyperfine states confined to one-dimensional geometries~\cite{PhysRevLett.105.190401,Pagano2014Nat.Phys.10.198}, where the internal states play the role of SU(4) internal degrees of freedom and the interaction strength can be tuned by confinement-induced resonances~\cite{PhysRevLett.81.938,giamarchi2004}. 
This makes the one-dimensional four-component Fermi gas an ideal benchmark system in which exact few-body quartet physics is known, allowing a stringent test of microscopic quartet BCS theory and Green's-function–based approaches.

In this study, we discuss an appropriate microscopic construction of the coherent state in the quartet BCS theory.
Applying it to a one-dimensional four-component Fermi gas with four-body interaction, we examine the validity of the quartet BCS frameworks with different approximations.
We show that the multiple-infinite-product ansatz successfully reproduces the exact four-body result while the single-infinite-product one gives a different renormalization condition.
As a step forward, we demonstrate that the self-consistent equation for the quartet superfluid order parameter obtained from the quartet BCS theory with the multiple-infinite-product ansatz is consistent with the equation derived by the generalized imaginary-time Nambu-Gor'kov formalism.
Moreover, we show the evolution of the single-particle spectral function obtained by a fixed-density, self-consistent calculation of the quartet superfluid order parameter at different interaction strengths.

This paper is organized as follows.
The Hamiltonian of the system to be investigated in this work is introduced in Sec.~\ref{sec:II}.
In Sec.~\ref{sec:III}, we discuss the exact four-body result to serve as a benchmark.
We then testify the validity of single- and multiple-infinite-product ansatzes in Sec.~\ref{sec:IVA} and Sec.~\ref{sec:IVB}, respectively.
In Sec.~\ref{sec:V}, we derive the gap equation via the generalized imaginary-time Nambu-Gor'kov formalism, and show the consistency with the quartet BCS theory.
In Sec.~\ref{sec:VI}, the self-consistent calculation of quartet BCS variational equations across the crossover is performed at a fixed density.
We track the evolution of full spectral function $A(k,\omega)$ across the crossover from the quartet superfluid to tetramer gas.
In addition, we also compare the results of single- and multiple-infinite-product ansatzes.
Finally, we summarize this paper in Sec.~\ref{sec:VII}.
In the following, we take $\hbar=c=k_{\rm B}=1$.

\section{Hamiltonian}\label{sec:II}

In order to testify the quartet Bardeen-Cooper-Schrieffer (BCS) theory, we consider the one-dimensional four-component Fermi gas.
For simplicity, we assume that only the four-body interaction survives, which can be described by the Hamiltonian as
\begin{align}
\label{hamiltonian}
H=\,&
\sum_{k} 
\left(
\xi_{k,a}a_{k}^\dag a_k
+
\xi_{k,b}b_{k}^\dag b_k
+
\xi_{k,c}c_{k}^\dag c_k
+
\xi_{k,d}d_{k}^\dag d_k
\right)\nonumber\\
& +\frac{g_0}{L^3}\sum_{k_i,k_i'}
a_{k_1}^\dag
b_{k_2}^\dag
c_{k_3}^\dag
d_{k_4}^\dag
d_{k_4'}
c_{k_3'}
b_{k_2'}
a_{k_1'} 
\nonumber\\
&\quad \times 
\delta_{k_1+k_2+k_3+k_4, k_1'+k_2'+k_3'+k_4'},
\end{align}
where $ \xi_{k, i} = {k^2}/{(2m_i)}-\mu_i$ is the single-particle energy with the atomic mass $m_i$ and chemical potential $\mu_i$.
For the sake of studying strong quartet correlations associated with four-body clusters, we neglect the two- and three-body interactions and the interactions among identical fermions.
The qualitative consequences of restoring the two-body interaction are discussed in Sec.~\ref{sec:VII}.
In addition, we consider the mass-balanced and population-balanced system as $m_i\equiv m$ and  $\mu_i\equiv \mu$.
The quartet creation operator with zero center-of-mass momentum is introduced as
\begin{align}
\label{eq:alpha_creation}
\alpha^\dagger_{p_1,p_2,p_3}
=\,&
a_{p_1}^\dag
b_{p_2}^\dag
c_{p_3}^\dag
d_{-p_1-p_2-p_3}^\dag
.
\end{align}
Hereafter, we denote $g_0/L^3$ as $g$ for the simplicity of the expressions.

\section{Exact four-body problem}\label{sec:III}

As a benchmark, we consider the four-body scattering problem in vacuum at zero center-of-mass momentum.
For the system described in Sec.~\ref{sec:II}, the four-body $T$ matrix with zero center-of-mass momentum frame satisfies
\begin{align}
    &T_4(\omega)
    =\left[\frac{1}{g}-
    \sum_{k_1,k_2,k_3}
    \frac{1}{\omega_+-\varepsilon_{k_1}-\varepsilon_{k_2}-\varepsilon_{k_3}-\varepsilon_{-(k_1+k_2+k_3)}}
    \right]^{-1},
\end{align}
where $\varepsilon_{k}=k^2/(2m)$ and $\omega^+=\omega+i0^+$.
After performing momentum integral with the momentum cutoff $\sqrt{k_1^2+k_2^2+k_3^2} \leq \Lambda$, one has
\begin{align}\label{eq6}
    \frac{4\pi a}{m}=T_4(0)=g_0\left(1+g_0\frac{m\Lambda}{3\sqrt{3}\pi}\right)^{-1}.
\end{align}
It is seen that the short-range four-body interaction in one dimension is characterized by the scattering length $a$ exactly in the same way that it characterizes the two-body one in three dimensions, which is normalized as
\begin{align}\label{eq2}
    -\frac{m}{4\pi a}=-\frac{1}{g_0}+\frac{1}{g_c},
\end{align}
where $g_c =-3\sqrt{3}\pi/
(m\Lambda)$ is the unitary coupling~\cite{Nishida2010Phys.Rev.A82.043606}.

The four-body binding energies $E_{\rm b}$ can be further obtained from 
\begin{widetext}
\begin{align}
\label{eq:6}
&[T_4(\omega=-E_{\rm b})]^{-1}\equiv
\left[\frac{m}{4\pi a}+
    \sum_{k_1,k_2,k_3}
    \left(\frac{1}{E_{\rm b}+\varepsilon_{k_1}+\varepsilon_{k_2}+\varepsilon_{k_3}+\varepsilon_{-(k_1+k_2+k_3)}}
    -\frac{1}{\varepsilon_{k_1}+\varepsilon_{k_2}+\varepsilon_{k_3}+\varepsilon_{-(k_1+k_2+k_3)}}
    \right)
    \right]=0,
\end{align}
\end{widetext}
which gives
\begin{align}
\label{eq:7}
E_{\rm b}=\frac{1}{2ma^2}.
\end{align}

\section{quartet BCS theory}\label{sec:IV}

\subsection{Single-infinite-product ansatz}\label{sec:IVA}

According to Refs.~\cite{Guo2022Phys.Rev.C105.024317, Guo2022Phys.Rev.Research4.023152}, the quartet BCS variational trial wave function is given by
\begin{align}
\label{eq:9}
    \left|\Psi^{\rm SIP}_{\rm QBCS}\right\rangle
    =\prod_{p}\left(u_{p}+v_{p}\alpha_p^\dag\right)\left|0\right\rangle,
\end{align}
with the normalization condition,
\begin{align}
    u_p^2+v_p^2=1.
\end{align}
Equation~\eqref{eq:9} is written in terms of an infinite product of $u_p+v_p\alpha_p^\dag$ due to the practical difficulty of handling multiple infinite products even in a numerical way. 
In this regard, we call the single-infinite-product (SIP) ansatz, which involves a BCS-like quartet operator with zero center-of-mass momentum and restricted internal structure $\{\pm p\}$ as
\begin{align}
    \alpha^{\dag}_p
    \equiv
\frac{1}{\sqrt{3}}
    \left(a_{p}^\dag
    b_{p}^\dag
    c_{-p}^\dag
    d_{-p}^\dag
    +
    a_{p}^\dag
    b_{-p}^\dag
    c_{p}^\dag
    d_{-p}^\dag
    +
    a_{p}^\dag
    b_{-p}^\dag
    c_{-p}^\dag
    d_{p}^\dag
    \right).
\end{align}
In the SIP ansatz, we restrict the quartet correlations to the BCS-like factorizable channel in which the four fermions occupy momenta $\pm p$.
Accordingly, the four-body interaction in Sec.~II is projected onto this restricted quartet subspace, leading to the effective Hamiltonian
\begin{align}
H_{\rm SIP}=\,&
\sum_{k} 
\left(
\xi_{k,a}a_{k}^\dag a_k
+
\xi_{k,b}b_{k}^\dag b_k
+
\xi_{k,c}c_{k}^\dag c_k
+
\xi_{k,d}d_{k}^\dag d_k
\right)\nonumber\\
&+g_{\rm SIP}\sum_{p,p'}
 \alpha^{\dag}_p
 \alpha_{p'},
\end{align}
where according to the projection one has
\begin{align}
    g_{\rm SIP}=3g=\frac{3g_0}{L^3}.
\end{align}
This truncation is the four-body analog of the Cooper-channel projection underlying the conventional BCS theory.
Based on that, the $T$ matrix matching with the momentum integral in the SIP ansatz reads
\begin{align}
    T_{\rm SIP}(\omega)
    =\left[\frac{1}{g_{\rm SIP}}-
    \sum_{k}
    \frac{1}{\omega_+-\varepsilon_{k}-\varepsilon_{k}-\varepsilon_{-k}-\varepsilon_{-k}}
    \right]^{-1}.
\end{align}
At $\omega=0$, we obtain
\begin{align}
    \sum_{k}
    \frac{1}{\omega_+-\varepsilon_{k}-\varepsilon_{k}-\varepsilon_{-k}-\varepsilon_{-k}}
    =-\frac{mL}{2\pi}\int_0^\Lambda\frac{{d}k}{k^2},
\end{align}
which indicates the IR divergence.
In the meanwhile, as for the four-body binding energy, one has
\begin{align}
&[T_{\rm SIP}(\omega=-E_{\rm b}^{\rm SIP})]^{-1}
=0,
\end{align}
which gives
\begin{align}
    1+g_{\rm SIP}L\frac{m}{4}\frac{1}{\sqrt{mE_{\rm b}^{\rm SIP}/2}}=0,
\end{align}
leading to
\begin{align}
    E_{\rm b}^{\rm SIP}=\frac{mg_{\rm SIP}^2L^2}{8}\equiv\frac{9mg_0^2}{8L^4}.
\end{align}
Consequently, as for the SIP ansatz in the quartet BCS theory, to avoid the infrared pathology of the zero-energy limit in one dimension, one needs to perform the renormalization at a finite negative energy $\omega=-E_{\rm b}^{\rm SIP}$, where the propagator is free of both IR and UV divergences.
In this sense, the renormalization condition of the coupling strength and its relation to $E_{\rm b}$ are different from the exact four-body result, due to the smaller Hilbert space of SIP. 
Practically, one can employ the interaction strength such that $E_{\rm b}$ matches the desired value. 
However, its validity and the relation to the exact four-body problem become obscure.

The quartet BCS variational equations under the SIP ansatz, namely, the gap and density equations are given as
\begin{align}
\rho
=\,&4\sum_k
\frac12\!\left(1-\frac{2\xi_k}{\sqrt{4\xi_k^2+\Delta_{\rm SIP}^2}}\right),\\
0=\,& 1+g_{\rm SIP}
\sum_k
\frac{1}{\sqrt{\xi_k^2+\Delta_{\rm SIP}^2/4}},
\end{align}
respectively~\cite{Guo2022Phys.Rev.C105.024317, Guo2022Phys.Rev.Research4.023152}.

\subsection{Multiple-infinite-product ansatz}\label{sec:IVB}

Based on the discussion given in Sec.~\ref{sec:IVA}, in order to fully reproduce the exact four-body problem investigated in Sec.~\ref{sec:III}, we recover the multiple-infinite-product (MIP) ansatz as derived in Refs.~\cite{Guo2022Phys.Rev.C105.024317,Guo2022Phys.Rev.Research4.023152}, which reads
\begin{align}\label{eqmip}
     \left|\Psi^{\rm MIP}_{\rm QBCS}\right\rangle
    =\prod_{p_1,p_2,p_3}
    \left(u_{p_1,p_2,p_3}+
    v_{p_1,p_2,p_3}
    \alpha_{p_1,p_2,p_3}^\dag \right)
     \left|0\right\rangle.
\end{align}
The normalization condition $\langle \Psi^{\rm MIP}_{\rm QBCS}| \Psi^{\rm MIP}_{\rm QBCS}\rangle=1$ leads to
\begin{align}
    |u_{p_1,p_2,p_3}|^2
    +|v_{p_1,p_2,p_3}|^2=1.
\end{align}
By applying the MIP ansatz on the Hamiltonian~\eqref{hamiltonian}, one has the expectation value of the kinetic part as
\begin{align}
    &\langle\Psi_{\rm QBCS}^{\rm MIP}|H_0|\Psi_{\rm QBCS}^{\rm MIP}\rangle\nonumber\\
    &\quad\quad=\,\sum_{k}\left[\prod_{k_2,k_3}
|v_{k,k_2,k_3}|^2\xi_{k}
+\prod_{k_1,k_3}
|v_{k_1,k,k_3}|^2\xi_{k}\right.\nonumber\\
&\left.\quad\qquad+\prod_{k_1,k_2}
|v_{k_1,k_2,k}|^2\xi_{k}
+\prod_{k_2,k_3}
|v_{k,k_2,k_3}|^2
\xi_{-k-k_2-k_3}
\right],
\end{align}
and the interaction term is evaluated as
\begin{align}
    &\langle\Psi_{\rm QBCS}^{\rm MIP}|V_4|\Psi_{\rm QBCS}^{\rm MIP}\rangle\nonumber\\
    =\,&g
    \sum_{k_1,k_2,k_3}
    \sum_{k_1'k_2',k_3'}
    v_{k_1,k_2,k_3}^*
    u_{k_1,k_2,k_3}
    v_{k_1',k_2',k_3'}
    u_{k_1',k_2',k_3'}^*.
\end{align}
By collecting all the terms, one obtains 
\begin{align}
    &\langle H\rangle
    =\,\sum_{k}\left[\prod_{k_2,k_3}
|v_{k,k_2,k_3}|^2\xi_{k}
+\prod_{k_1,k_3}
|v_{k_1,k,k_3}|^2\xi_{k}\right.\nonumber\\
&\quad\left.\qquad+\prod_{k_1,k_2}
|v_{k_1,k_2,k}|^2\xi_{k}
+\prod_{k_2,k_3}
|v_{k,k_2,k_3}|^2
\xi_{-k-k_2-k_3}
\right]\nonumber\\
    &\quad\quad\quad+g
    \sum_{k_1,k_2,k_3}
    \sum_{k_1'k_2',k_3'}
    v_{k_1,k_2,k_3}^*
    u_{k_1,k_2,k_3}
    v_{k_1',k_2',k_3'}
    u_{k_1',k_2',k_3'}^*.
\end{align}

From the normalization condition, with setting $u_{k_1,k_2,k_3}$ to be real, one obtains
\begin{align}
    \delta u_{k_1,k_2,k_3}=-\frac{1}{2u_{k_1,k_2,k_3}}v_{k_1,k_2,k_3}\delta v_{k_1,k_2,k_3}^*.
\end{align}
In analogy with the BCS theory, the quartet superfluid order parameter is defined as
\begin{align}\label{eq25}
    \Delta=-g\sum_{k_1',k_2',k_3'}    
u
_{k_1',k_2',k_3'}
    v_{k_1',k_2',k_3'}.
\end{align}
By taking the variation of the expectation value of the Hamiltonian, one obtains
\begin{align}
    \delta \langle H\rangle =\,&
v_{k_1,k_2,k_3}\delta v_{k_1,k_2,k_3}^*
    \left(
    \xi_{k_1}+\xi_{k_2}+\xi_{k_3}
    +\xi_{-k_1-k_2-k_3}
    \right)\nonumber\\
    &\times\prod_{p_2\neq k_2,p_3\neq k_3}
|v_{k_1,p_2,p_3}|^2
    \nonumber\\
    &-
    \left(
    u_{k_1,k_2,k_3}\delta v_{k_1,k_2,k_3}^*
+
v_{k_1,k_2,k_3}^*
   \delta u_{k_1,k_2,k_3}
    \right)
    \Delta\nonumber\\
    &-
    v_{k_1,k_2,k_3}
    \delta u_{k_1,k_2,k_3}
    \Delta^*.
\end{align}
Furthermore, using
\begin{align}
\frac{\delta}{\delta u_{k_1,k_2,k_3}}\langle H\rangle-
    \frac{2u_{k_1,k_2,k_3}}{v_{k_1,k_2,k_3}}\frac{\delta}{\delta v_{k_1,k_2,k_3}^\ast}\langle H\rangle=0,
\end{align}
one obtains
\begin{align}\label{eq28}
    v_{k_1,k_2,k_3}=
    \frac{u_{k_1,k_2,k_3}\Delta}{B_{k_1,k_2,k_3}+\mathcal{E}_k\prod_{p_2\neq k_2,p_3\neq k_3}
|v_{k_1,p_2,p_3}|^2}
    ,
\end{align}
where
\begin{align}
    B_{k_1,k_2,k_3}
    =\,&\frac{1}{2u_{k_1,k_2,k_3}}\left(v^\ast_{k_1,k_2,k_3}\Delta+v_{k_1,k_2,k_3}\Delta^\ast\right)\nonumber\\
    =\,&\frac{1}{u_{k_1,k_2,k_3}}\operatorname{Re}\left(v^\ast_{k_1,k_2,k_3}\Delta\right)
\end{align}
and
\begin{align}
    \mathcal{E}_k=\xi_{k_1}+\xi_{k_2}+\xi_{k_3}+\xi_{-k_1-k_2-k_3}
\end{align}
are introduced, respectively.
In the present population-balanced system, the total number density equation reads
\begin{align}
    N=4\sum_{k}\prod_{k_2,k_3}|v_{k,k_2,k_3}|^2.
\end{align}
Equation~\eqref{eq28} can be recast into
\begin{align}
    B_{k_1,k_2,k_3}=\frac{\Delta^2}{B_{k_1,k_2,k_3}+\mathcal{E}_k\prod_{p_2\neq k_2,p_3\neq k_3}
|v_{k_1,p_2,p_3}|^2},
\end{align}
namely,
\begin{align}
    B_{k_1,k_2,k_3}^2+\mathcal{E}_k 
    \mathcal{G}_{k_1,k_2,k_3}
    B_{k_1,k_2,k_3}-\Delta^2=0,
\end{align}
where 
\begin{align}
    \mathcal{F}_{k}=\prod_{p_2,p_3}|v_{k,p_2,p_3}|^2
\end{align}
and
\begin{align}
    \prod_{p_2\neq k_2,p_3\neq k_3}
|v_{k_1,p_2,p_3}|^2
=\frac{\mathcal{F}_{k_1}}{|v_{k_1,k_2,k_3}|^2}\equiv \mathcal{G}_{k_1,k_2,k_3}
\end{align}
are introduced for convenience.
Under the ambiguity of the sign originating from the broken U($1$) symmetry, we adopt
\begin{align}
    B_{k_1,k_2,k_3}=-\frac{1}{2}\mathcal{E}_k\mathcal{G}_{k_1,k_2,k_3}+\sqrt{\frac{1}{4}\mathcal{E}_k^2\mathcal{G}_{k_1,k_2,k_3}^2+\Delta^2}.
\end{align}
Consequently, one has
\begin{align}
    u_{k_1,k_2,k_3}^2&=\frac{1}{2}\left(1+\frac{\frac{1}{2}\mathcal{E}_k\mathcal{G}_{k_1,k_2,k_3}}{\sqrt{\frac{1}{4}\mathcal{E}_k^2\mathcal{G}_{k_1,k_2,k_3}^2+\Delta^2}}\right),\\
     v_{k_1,k_2,k_3}^2&=\frac{1}{2}\left(1-\frac{\frac{1}{2}\mathcal{E}_k\mathcal{G}_{k_1,k_2,k_3}}{\sqrt{\frac{1}{4}\mathcal{E}_k^2\mathcal{G}_{k_1,k_2,k_3}^2+\Delta^2}}\right).
\end{align}
It can be rewritten as
\begin{align}
    \mathcal{G}_{p_1,p_2,p_3}=
    \prod_{k_2\neq p_2,k_3\neq p_3}
    \left[
    \frac{1}{2}
    \left(1-\frac{\frac{1}{2}\mathcal{E}_k\mathcal{G}_{k_1,k_2,k_3}}{\sqrt{\frac{1}{4}\mathcal{E}_k^2\mathcal{G}_{k_1,k_2,k_3}^2+\Delta^2}}\right)
    \right].
\end{align}
\begin{widetext}
Further combined with Eq.~\eqref{eq25}, the quartet BCS gap equation reads
\begin{align}\label{eqgap}
    \frac{m}{4\pi a}
    +\sum_{k_1,k_2,k_3}\left[\frac{1}{\sqrt{\mathcal{E}_k^2\mathcal{G}_{k_1,k_2,k_3}^2+4\Delta^2}}
    -\frac{1}{\varepsilon_{k_1}+\varepsilon_{k_2}+\varepsilon_{k_3}+\varepsilon_{-k_1-k_2-k_3}}
    \right]=0.
\end{align}
\end{widetext}
One can notice that Eq.~\eqref{eqgap} is similar to Eq.~\eqref{eq:6}.
This result indicates that $\sqrt{\mathcal{E}_k^2\mathcal{G}_{k_1,k_2,k_3}^2+4\Delta^2}\simeq E_{\rm b}+\varepsilon_{k_1}+\varepsilon_{k_2}+\varepsilon_{k_3}+\varepsilon_{-(k_1+k_2+k_3)}$ in the dilute limit as in the case of the BCS-Eagles-Leggett theory~\cite{Ohashi2020Prog.Part.Nucl.Phys.111.103739}. This condition is satisfied when $\mu\simeq -E_{\rm b}/4$ and $|\Delta|/|\mu|\ll1$.
In this regard, it is also seen that the MIP ansatz successfully recovers the exact four-body results in the dilute limit, with the four-body binding energy $E_{\rm b}^{\rm MIP}=1/(2ma^2)$ being consistent with Eq.~\eqref{eq:7}.

\section{Generalized Nambu-Gor'kov formalism}\label{sec:V}

In Ref.~\cite{Sogo2010Phys.Rev.C81.064310}, the Nambu-Gor'kov formalism has been generalized for quarteting.
In this section, we demonstrate that the gap equation obtained from the Nambu-Gor'kov formalism is consistent with the one derived from the MIP ansatz of quartet BCS theory given by Eq.~\eqref{eqgap}.

In order to derive the gap equation within the Nambu-Gor'kov formalism, it is convenient to use an imaginary-time formalism with imaginary time $\tau$.
On the basis of Ref.~\cite{Sogo2010Phys.Rev.C81.064310}, the generalized Nambu-Gor'kov spinors are given by
\begin{align}
    \bm{A}_{k_i}=\left(
    \begin{array}{c}
         a_{k_1}  \\
         d_{-k_1-k_2-k_3}^\dag c_{k_3}^\dag b_{k_2}^\dag 
    \end{array}
    \right)
\end{align}
and
\begin{align}
    \bm{A}_{k_i}^\dag=
    \left(
    \begin{array}{cc}
       a_{k_1}^\dag  &  b_{k_2}c_{k_3}d_{-k_1-k_2-k_3}
    \end{array}
    \right).
\end{align}
\begin{widetext}
The Nambu-Gor'kov Green's function thus reads
\begin{align}
    &\langle T_\tau[\bm{A}_{k_i}(\tau)\bm{A}_{k_i}^\dag(0)]\rangle\nonumber\\
    =\,&
    \left(
    \begin{array}{cc}
       \langle T_\tau [a_{k_1}(\tau)a_{k_1}^\dag(0)]\rangle  & \langle T_\tau[a_{k_1}(\tau)b_{k_2}(0)c_{k_3}(0)d_{-k_1-k_2-k_3}(0)]\rangle  \\
        \langle T_\tau[d_{-k_1-k_2-k_3}^\dag(\tau)c_{k_3}^\dag(\tau)b_{k_2}^\dag(\tau)a_{k_1}^\dag(0)]\rangle & 
         \langle T_\tau[d_{-k_1-k_2-k_3}^\dag(\tau)c_{k_3}^\dag(\tau)b_{k_2}^\dag(\tau)
         b_{k_2}(0)c_{k_3}(0)d_{-k_1-k_2-k_3}(0)
         ]\rangle
    \end{array}
    \right)\cr
    \equiv\,& 
    \left(
    \begin{array}{cc}
      -G_{k_1}(\tau)   & -F_{k_i}(\tau) \\
       -F^\dag_{k_i}(\tau)  & -\mathcal{G}_{k_i}(-\tau)
    \end{array}
    \right).
\end{align}

The equation of motion of the Nambu-Gor'kov Green's function is obtained from
\begin{align}
    \frac{\partial}{\partial \tau}
     \langle T_\tau[\bm{A}_{k_i}(\tau)\bm{A}_{k_i}^\dag(0)]\rangle
     =\delta(\tau)\langle
     \{\bm{A}_{k_i},\bm{A}_{k_i}^\dag\}
     \rangle 
     +\langle T_\tau[\partial_\tau\bm{A}_{k_i}(\tau)\bm{A}_{k_i}^\dag(0)]\rangle.
\end{align}
With zero center-of-mass momentum, the matrix $\langle \{\bm{A}_{k_i},\bm{A}_{k'_i}^\dag\}\rangle$ is rewritten as
\begin{align}
    \langle \{\bm{A}_{k_i},\bm{A}_{k'_i}^\dag\}\rangle
    =\,&
    \left(
    \begin{array}{cc}
       \langle \{a_{k_1},a_{k'_1}^\dag\}\rangle  & 
       \langle \{a_{k_1},b_{k'_2}c_{k'_3}d_{-k'_1-k'_2-k'_3}\}\rangle \\
        \langle \{d^\dag_{-k_1-k_2-k_3}c^\dag_{k_3}b^\dag_{k_2},a_{k'_1}^\dag\}\rangle & 
         \langle \{d^\dag_{-k_1-k_2-k_3}c^\dag_{k_3}b^\dag_{k_2},b_{k'_2}c_{k'_3}d_{-k'_1-k'_2-k'_3}\}\rangle
    \end{array}
    \right)\cr
      =\,&
    \left(
    \begin{array}{cc}
       \delta_{k_1,k_1'}  & 
       0 \\
        0 & 
         \delta_{k_1,k_1'}\delta_{k_2,k_2'}\delta_{k_3,k_3'}\mathcal{N}_{k_2,k_3,-k_1-k_2-k_3}
    \end{array}
    \right),
\end{align}
where we introduce the notation 
\begin{align}
    \mathcal{N}_{k_1,k_2,k_3}=
    f_{k_1}f_{k_2}f_{k_3}+\bar{f}_{k_1}\bar{f}_{k_2}\bar{f}_{k_3},
\end{align}
with
\begin{align}
    f_{k}=\frac{1}{e^{\beta \xi_k}+1},\quad\bar{f}_k=1-f_k.
\end{align}
On the other hand, according to the Heisenberg equation of motion, one obtains
\begin{align}
    -\langle T_\tau[\partial_\tau a_{k_1}(\tau)a_{k_1}^\dag(0)]\rangle
    &=
    \langle T_\tau[[a_{k_1}(\tau),H]a_{k_1}^\dag(0)]\rangle
\end{align}
and
\begin{align}
    -\langle T_\tau[\partial_\tau \{d_{k_4}^\dag(\tau)c_{k_3}^\dag(\tau)b_{k_2}^\dag(\tau)\}a_{k_1}^\dag(0)]\rangle
    &=
    \langle T_\tau[[d_{k_4}^\dag(\tau)c_{k_3}^\dag(\tau)b_{k_2}^\dag(\tau),H]a_{k_1}^\dag(0)]\rangle.
\end{align}
The commutation relations are given by
\begin{align}
    [a_{k_1},H]
    &=\xi_{k_1,a}a_{k_1}+g\sum_{k_2,k_3,k_4,k_1',k_2',k_3',k_4'}
    \delta_{k_1+k_2+k_3+k_4,k_1'+k_2'+k_3'+k_4'}
    a_{k_1'}b_{k_2}^\dag c_{k_3}^\dag d_{k_4}^\dag
    d_{k_4'} c_{k_3'} b_{k_2'},
\end{align}
\begin{align}
    [d_{k_4}^\dag c_{k_3}^\dag b_{k_2}^\dag,H]
=\,&
    -(\xi_{k_4,d} +\xi_{k_3,c}+\xi_{k_2,b})
    d_{k_4}^\dag c_{k_3}^\dag b_{k_2}^\dag\cr
    &
    -g\sum_{k_i',p_i}
     \delta_{p_1+p_2+p_3+p_4,k_1'+k_2'+k_3'+k_4'}
    a_{k_1'}^\dag
    b_{k_2'}^\dag 
    c_{k_3'}^\dag 
    d_{k_4'}^\dag
    a_{p_1}
    \mathcal{N}_{p_2,p_3,p_4}
    \delta_{p_2,k_2}
    \delta_{p_3,k_3}
    \delta_{p_4,k_4}.
\end{align}
As a result, one has
\begin{align}
    &-\langle T_\tau[\partial_\tau a_{k_1}(\tau)a_{k_1}^\dag(0)]\rangle\nonumber\\
    =\,&
    \xi_{k_1,a}\langle T_\tau[a_{k_1}(\tau)a_{k_1}^\dag(0)]\rangle
     \cr
   &-
    g\sum_{k_2,k_3,k_4,{k_1'},k_2',k_3',k_4'}
    \delta_{k_1+k_2+k_3+k_4,k_1'+k_2'+k_3'+k_4'}
    \langle T_\tau[
    d_{k_4}^\dag(\tau)
    c_{k_3}^\dag(\tau)
    b_{k_2}^\dag(\tau)  
    a_{k_1}^\dag(0)
    a_{k_1'}(\tau)
    b_{k_2'}(\tau)
     c_{k_3'}(\tau)
     d_{k_4'}(\tau)
    ]\rangle\cr
    =\,&
    \xi_{k_1,a}\langle T_\tau[a_{k_1}(\tau)a_{k_1}^\dag(0)]\rangle
   -
    g\sum_{k_2,k_3,k_4,{k_1'},k_2',k_3',k_4'}
    \delta_{k_1+k_2+k_3+k_4,k_1'+k_2'+k_3'+k_4'}
  F_{k_i}^\dag(\tau)
    F_{k_i'}(0),
\end{align}
leading to the equation of motion of $G_{k_1}(\tau)$ as
\begin{align}
    \left[\partial_\tau +\xi_{k_1,a}\right]G_{k_1}(\tau)
    = -\delta(\tau)
    -\sum_{k_2,k_3,k_4}
    F_{k_i}^\dag(\tau)
    \left[g\sum_{k_i'}
    \delta_{k_1+k_2+k_3+k_4,k_1'+k_2'+k_3'+k_4'}
    F_{k_i'}(0)\right].
\end{align}
In a similar manner, one can obtain
\begin{align}
[-\partial_\tau +\xi_{k_4,d}+\xi_{k_3,c}+\xi_{k_2,b}]F_{k_i}^\dag(\tau)
         = \,&
         \mathcal{N}_{k_2,k_3,k_4}
         G_{k_1}(\tau)
         \left[g
         \sum_{k_i'}
     \delta_{k_1+k_2+k_3+k_4,k_1'+k_2'+k_3'+k_4'}
         F^\dag_{k_i'}(0)\right].
\end{align}
With the Fourier transformation to Matsubara frequency as
\begin{align}
    G_{k}(\tau)=\frac{1}{\beta}\sum_{n}G_{k}(i\omega_n)e^{-i\omega_n\tau},\quad
    F_{k}^\dag(\tau)=\frac{1}{\beta}\sum_{n}F_{k}^\dag(i\omega_n)e^{-i\omega_n\tau},
\end{align}
one obtains
\begin{align}
    \left[-i\omega_n +\xi_{k_1,a}\right]
    G_{k_1}(i\omega_n)
=-1
    -\Delta
    \sum_{k_2,k_3,k_4}
    F^\dag_{k_i}(i\omega_n)
\end{align}
and
\begin{align}
    [i\omega_n +\xi_{k_4,d}+\xi_{k_3,c}+\xi_{k_2,b}]
F^\dag_{k}(i\omega_n)
         = 
         \Delta^\ast
         \mathcal{N}_{k_2,k_3,k_4}
         G_{k_1}(i\omega_n),
\end{align}
where the quartet order parameter is defined as
\begin{align}\label{eq59}
    \Delta=\frac{1}{\beta}g\sum_{k_i'}
    \sum_n 
    \delta_{k_1+k_2+k_3+k_4,k_1'+k_2'+k_3'+k_4'}
    F_{k_i'}(i\omega_n).
\end{align}

From above equations and taking the zero center-of-mass momentum, one has
\begin{align}
    F_{k_i}^\dag(i\omega_n)=\Delta^*
    \mathcal{N}_{k_2,k_3,-(k_1+k_2+k_3)}
    \frac{G_{k_1}(i\omega_n)}{i\omega_n+\xi_{-(k_1+k_2+k_3),d}+\xi_{k_3,c}+\xi_{k_2,b}}.
\end{align}
Consequently, the quartet gap equation can be given as
\begin{align}\label{eq61}
    \Delta^*= \frac{g}{\beta}
    \sum_{n}
    \sum_{k_i'}
    \frac{\Delta^*\mathcal{N}_{k_2',k_3',k_4'}G_{k_1'}(i\omega_n)}{i\omega_n+\xi_{k_4',d}+\xi_{k_3',c}+\xi_{k_2',b}},
\end{align}
where
\begin{align}\label{eq62}
    G_{k_1}(i\omega_n)
    = \frac{1}{i\omega_n-\xi_{k_1,a}-|\Delta|^2
    \sum_{k_2,k_3}
    \frac{\mathcal{N}_{k_2,k_3,-(k_1+k_2+k_3)}}{i\omega_n +[\xi_{-(k_1+k_2+k_3),d}+\xi_{k_3,c}+\xi_{k_2,b}]}},
\end{align}
and the corresponding diagrammatic illustration of the quartet mean-field self-energy in the condensed phase is shown as Fig.~1.
Eq.~\eqref{eq62} indicates that the Green's function is dressed by the self-energy 
\begin{align}
    \Sigma_k(i\omega_n)=|\Delta|^2
    \sum_{k_2,k_3}
    \frac{\mathcal{N}_{k_2,k_3,-(k_1+k_2+k_3)}}{i\omega_n +[\xi_{-(k_1+k_2+k_3),d}+\xi_{k_3,c}+\xi_{k_2,b}]},
\end{align}
which is diagrammatically illustrated in Fig.~\ref{Fig:1}.
$\Sigma_k(i\omega_n)$ represents the one-particle-three-hole coupling due to the quartet correlations with $\Delta$.
Accordingly,  Eq.~\eqref{eq61} is rewritten as
\begin{align}\label{eq178}
    1=\frac{g}{\beta}
    \sum_{n}\sum_{k_i'}\frac{
    \mathcal{N}_{k_2',k_3',k_4'}}{\left(i\omega_n+\xi_{k_4',d}+\xi_{k_3',c}+\xi_{k_2',b}
    \right)
    \left(
    i\omega_n-\xi_{k_1,a}-|\Delta|^2
    \sum_{k_2,k_3}
    \frac{\mathcal{N}_{k_2,k_3,-(k_1+k_2+k_3)}}{i\omega_n +[\xi_{-(k_1+k_2+k_3),d}+\xi_{k_3,c}+\xi_{k_2,b}]}
    \right)
    }.
\end{align}

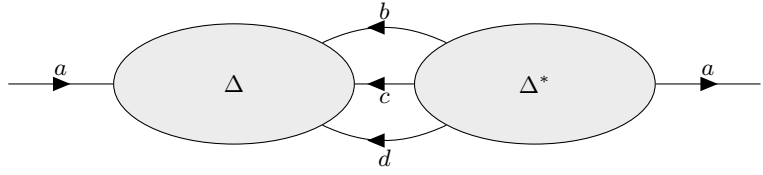
\begin{figure}
\begin{tikzpicture}
\begin{feynman}
\vertex[draw, fill=gray!15, ellipse,
        minimum width=3.2cm, minimum height=1.6cm]
        (vL) at (-2,0) {\(\Delta\)};
\vertex[draw, fill=gray!15, ellipse,
        minimum width=3.2cm, minimum height=1.6cm]
        (vR) at ( 2,0) {\(\Delta^\ast\)};
\vertex (ain)  at (-5,0);
\vertex (aout) at ( 5,0);
\diagram*{
  (ain)  -- [fermion, edge label=\(a\)] (vL),
  (vR)   -- [fermion, edge label=\(a\)] (aout),
};
\diagram*{
  (vL) -- [anti fermion, bend left=25,  edge label=\(b\)] (vR),
  (vL) -- [anti fermion,               edge label'=\(c\)] (vR),
  (vL) -- [anti fermion, bend right=25, edge label'=\(d\)] (vR),
};
\end{feynman}
\end{tikzpicture}
\caption{
The diagrammatic illustration of the quartet mean-field self-energy in the condensed phase~\eqref{eq62}.
The blobs denote the quartet order parameter, and the internal lines represent the effective three-particle propagator~\cite{Tajima2022Phys.Rev.Research4.L012021}.}
\end{figure}

The Matsubara summation is performed by replacing it with the contour integral enclosing the pole of $-\beta/(e^{\beta z}+1)$ at $z=i\omega_n$,
\begin{align}
    1&=\frac{g}{\beta}
    \sum_{n}\sum_{k_i'}\frac{
    \mathcal{N}_{k_2',k_3',k_4'}}{\left(i\omega_n+\xi_{k_4',d}+\xi_{k_3',c}+\xi_{k_2',b}
    \right)
    \left(
    i\omega_n-\xi_{k_1,a}-|\Delta|^2
    \sum_{k_2,k_3}
    \frac{1}{i\omega_n +[\xi_{-(k_1+k_2+k_3),d}+\xi_{k_3,c}+\xi_{k_2,b}]}
    \right)
    }
    \cr
    &=  -g
    \sum_{k_i'}
    \oint\frac{dz}{2\pi i}\frac{1}{e^{\beta z}+1}
    \sum_{k_i'}
     \frac{
    \mathcal{N}_{k_2',k_3',k_4'}
    }{\left(z+\xi_{k_4',d}+\xi_{k_3',c}+\xi_{k_2',b}
    \right)
    \left(
    z-\xi_{k_1,a}-|\Delta|^2  \sum_{k_2,k_3}\frac{1}{z+\xi_{-(k_1+k_2+k_3),d}+\xi_{k_3,c}+\xi_{k_2,b}}\right)}.
\end{align}
In comparison with the quartet BCS theory, here we further assume that the momentum transfers are sufficiently small as $k_i\simeq k_i'$ in Eq.~\eqref{eq178}.  
Then, one obtains
\begin{align}
    1&\simeq
    -g\sum_{k_1,k_2,k_3}
    \mathcal{N}_{k_2,k_3,-(k_1+k_2+k_3)}
    \left[\frac{f(E_{k_i,-})-f(E_{k_i,+})}{
\sqrt{(\xi_{k_1,a}+\xi_{-(k_1+k_2+k_3),d}+\xi_{k_3,c}+\xi_{k_2,b})^2+4|\Delta|^2\mathcal{N}_{k_2,k_3,-(k_1+k_2+k_3)}}
    }
    \right],
\end{align}
where $f(x)=1/(e^{x/T}+1)$ is the Fermi-Dirac distribution function and
\begin{align}
    E_{k_i,\pm}
    =\frac{\xi_{k_1,a}-[\xi_{-(k_1+k_2+k_3),d}+\xi_{k_3,c}+\xi_{k_2,b}]\pm\sqrt{[\xi_{k_1,a}+\xi_{-(k_1+k_2+k_3),d}+\xi_{k_3,c}+\xi_{k_2,b}]^2+4|\Delta|^2\mathcal{N}_{k_2,k_3,-(k_1+k_2+k_3)}}}{2}.
\end{align}
\end{widetext}

At $T=0$, we can also assume that the region near the Fermi surface $\xi_{k_i,a(bcd)}\simeq 0$ is relevant and only the lower branch is fully occupied as $f(E_{k_i,-})\simeq 1$, leading to 
\begin{align}\label{eqgapgo}
    1 +g\sum_{k_1, k_2,k_3}
    \frac{    \mathcal{N}_{k_2,k_3,-(k_1+k_2+k_3)}
}{
\sqrt{\mathcal{E}_k^2+4|\Delta|^2\mathcal{N}_{k_2,k_3,-(k_1+k_2+k_3)}}
    }=0.
\end{align}
In the dilute limit, one also has $\mathcal{N}_{k_2,k_3,-(k_1+k_2+k_3)}\simeq1$ by taking $f_{k_i}\simeq 0$.
Together with the normalization relation~\eqref{eq6}, it could be seen that the gap equation~\eqref{eqgapgo} obtained via the generalized imaginary-time Nambu-Gor'kov formalism is equivalent to the gap equation in quartet BCS theory with the MIP ansatz as Eq.~\eqref{eqgap}.

\section{Numerical demonstration of quartet superfluid description}
\label{sec:VI}

To gain physical insights of quartet superfluid state described by the generalized Nambu-Gor'kov formalism and the quartet BCS theory, we present the numerical demonstration of the single-particle excitation and the quartet order parameters.
In this section, we set $k_{\rm F}=E_{\rm F}=1$, the system size $L$ to unity, and the momentum cutoff $\Lambda=12$. 
The number density equation can be obtained as
\begin{align}
\label{eq:68}
    \rho=4\sum_{\bm{k}}\int_{-\infty}^{\infty} d\omega f(\omega)A_{k}(\omega),
\end{align}
where
\begin{align}
    A_{k}(\omega)&=-\frac{1}{\pi}{\rm Im}\,G_{k}(\omega)
\end{align}
is the single-particle spectral function with the retarded Green's function
\begin{align}
    G_{k}(\omega)&=\frac{1}{\omega+i\delta-\xi_{k}-\sum_{k_2,k_3}\frac{\Delta^2}{\omega+i\delta+\xi_{k_2}+\xi_{k_3}+\xi_{-k-k_2-k_3}}}.
\end{align}
We obtain the normalized chemical potential $\mu$ and quartet order parameter $\Delta$ by numerically solving Eqs.~\eqref{eqgapgo} and \eqref{eq:68}
at a fixed density $\rho=4/\pi$.
Moreover, in Eq.~\eqref{eqgapgo}, we assume
$\mathcal{N}_{k_1,k_2,k_3}\simeq 1$.
While this approximation is reliable at low densities and leads to the overestimation of quartet correlations at higher densities, it is sufficient for our qualitative demonstration. Note that our results are confirmed to be qualitatively unchanged for different cutoffs.

\begin{figure}
  \includegraphics[width=0.45\textwidth]{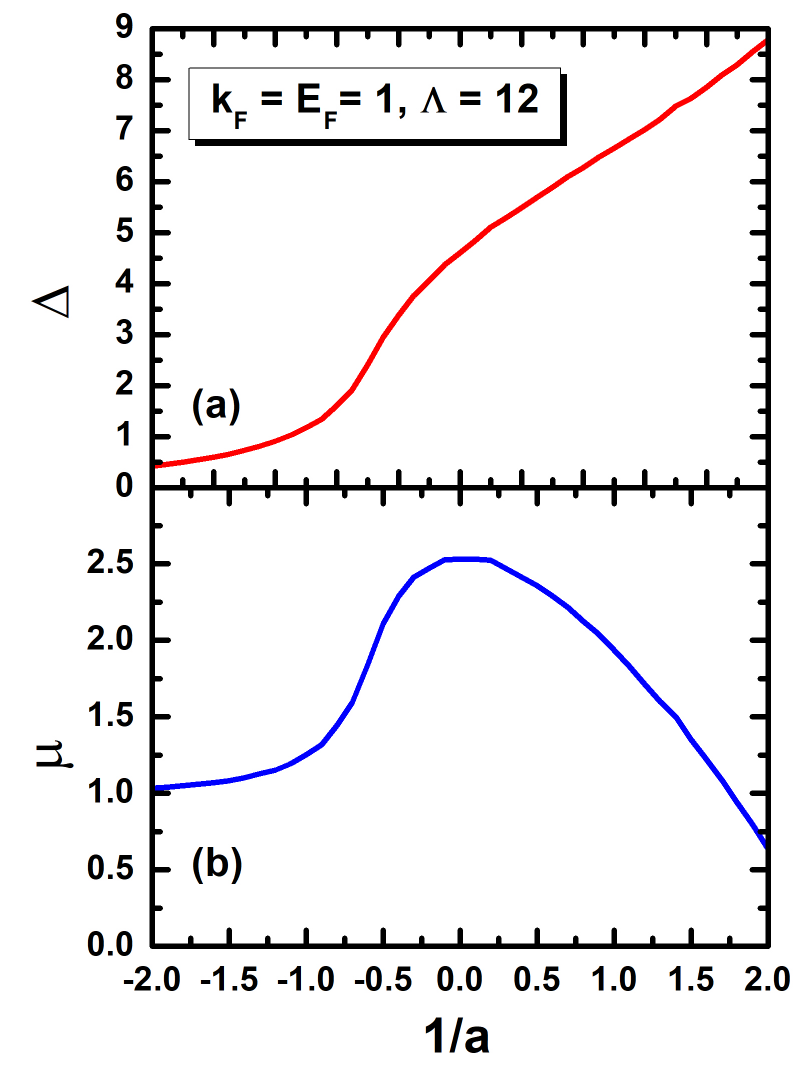}
  \caption{(a)~The order parameter $\Delta$ as a function of the inverse scattering length $1/a$.
  (b)~The chemical potential $\mu$ as a function of the inverse of scattering length $1/a$.
  The density is fixed at $4/\pi$.
  }\label{Fig:1}
\end{figure}

In Fig.~\ref{Fig:1} we show the self-consistent solutions of $\Delta$ and $\mu$ as functions of the inverse scattering length $1/a$ at a fixed density $\rho = {4}/{\pi}$.
While $\Delta$ monotonically increases with $1/a$ as in the case of the BCS-BEC crossover~\cite{Strinati2018Phys.Rep.738.1--76,Ohashi2020Prog.Part.Nucl.Phys.111.103739}, $\mu$ shows a maximum around $1/a=0$ and turn to decrease at larger $1/a$. The peaked behavior of $\mu$ is reminicent of the temperature dependence of $\mu$ in a one-dimensional free Fermi gas~\cite{johnston2021thermodynamicsnonrelativisticfreeelectronfermi}. 
Thus, it can be a specific feature of the one-dimensional system rather than that of the quartet superfluid.
Although the present result of $\Delta$ is basically different from the excitation gap reported in Ref.~\cite{Nishida2010Phys.Rev.A82.043606},
the increasing behavior of $\Delta$ is qualitatively consistent with the weak-coupling result based on the effective field theory~\cite{Nishida2010Phys.Rev.A82.043606}.

For the numerical result of $\mu$, our numerical demonstration is larger than the Bertsch parameter at unitarity ($1/a=0$) defined in terms of the chemical potential, which is given as $0.37$~\cite{PhysRevLett.109.250403}.
In contrast, $\mu$ in the present calculation can include a sizable static Brueckner-Hartree-Fock (BHF) contribution, which is generally negative for attractive multi-body interactions~\cite{PhysRevA.85.012701,PhysRevA.95.043625,PhysRevC.109.055203}.
Suppose that $\mu$ used in the quartet BCS theory is already shifted by $\Sigma_{\rm BHF}$, prior to incorporating explicit quartet correlations.
Consequently, $\mu$ obtained from the density equation and shown in Fig.~\ref{Fig:1}(b) corresponds to a shifted chemical potential compared to the quantum Monte Carlo calculation, that is,
\begin{align}
\mu = \mu_{\rm bare} - \Sigma_{\mathrm{BHF}}>\mu_{\rm bare},
\end{align}
where $\mu_{\rm bare}$ is the bare chemical potential.
In this regard, the discrepancy between our result of $\mu$ at $1/a=0$ in Fig.~\ref{Fig:1}(b) and the Bertsch parameter $\mu_{\rm bare}\simeq 0.37$ in Ref.~\cite{PhysRevLett.109.250403} can thus be attributed to this static self-energy shift.
From this viewpoint, the qualitative evolution across the crossover remains physically
consistent, and the large positivity of $\mu$ in Fig.~\ref{Fig:1}(b) could be interpreted as a consequence of the sizable BHF energy shift inherent in the present approximation.
We note that, while $\mu$ is positive in the crossover region of Fig.~\ref{Fig:1},
eventually $\mu$ turns to be negative at stronger couplings (e.g., at $1/a=3$, we obtain $\mu\simeq -1.33$).

\begin{figure}
  \includegraphics[width=0.45\textwidth]{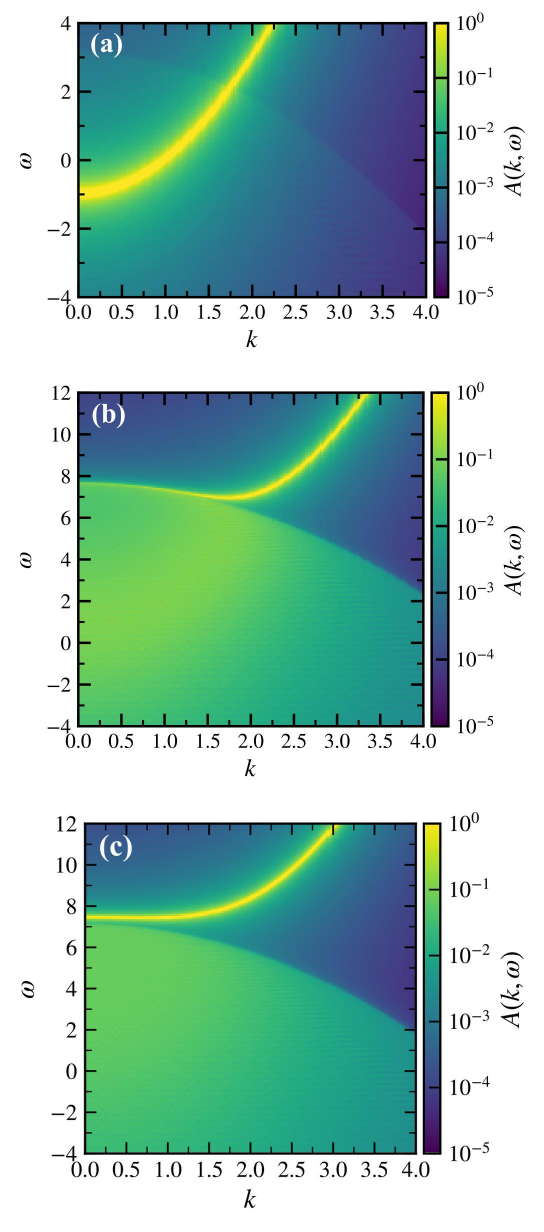}
  \caption{
  Contour plot of the spectral function $A_k(\omega)$ at
  (a) $1/a=-2.0$, (b) $1/a=0.0$, and (c) $1/a=0.5$, respectively.
  Here $k_{\rm F}=E_{\rm F}=1$ and the momentum cutoff is $\Lambda=12$.
The color bar is in the logarithmic scale, representing the intensity of $A(k,\omega)$.
  }
  \label{Fig:2}
\end{figure}

Fig.~\ref{Fig:2} shows the contour plots of the single-particle spectral function $A(k,\omega)$ for different interaction strengths $1/a$.
In the weak-coupling side ($1/a<0$), Fig.~\ref{Fig:2}(a) displays a clear particle-like quasiparticle branch at positive frequencies.
In contrast, the negative-frequency region is broadened by a continuum originating from three-particle excitations.
This is a signature of the formation of Cooper quartets where the annihilation of three particles (i.e., the creation of three holes) is needed to excite a single particle from a four-body object.
We note that in the present theory, the self-energy is dominated by three-hole intermediate states, which strongly breaks particle–hole symmetry and leads to a single dominant dispersive feature in the spectral function.

At $1/a=0$ shown in Fig.~\ref{Fig:2}(b), while the spectral weight at positive frequencies is still present, the quasiparticle-like branch is strongly broadened and no longer well separated from the incoherent background.
In particular, the spectral function shows a pronounced asymmetry between positive and negative frequencies, reflecting the dominance of three-hole intermediate states in the self-energy.
This behavior signals the onset of a strongly correlated regime dominated by incoherent scattering processes associated with four-body correlations.

In the strong-coupling regime shown in Fig.~\ref{Fig:2}(c) at $1/a=0.5$, the dominant low-energy fermionic quasiparticles are absent.
In this regime, the single-particle spectral function is no longer dominated by long-lived fermionic quasiparticles at low energies.
Instead, a large fraction of the spectral weight is transferred to incoherent continua associated with strongly bound quartets and quartet-breaking excitations.
A sharp dispersive feature appears at high positive frequencies.
This branch originates from the single-particle excitation from a bound quartet state by removing a single fermion, i.e., a molecular pole in the three-hole self-energy channel.
At unitarity, although a dispersive feature can also be identified on the positive-frequency side, its spectral weight remains broadly distributed and extends down to low frequencies, consistent with strongly damped fermionic excitations.
In the strong-coupling side, by contrast, the spectral response is dominated by a sharp high-energy peak whose small-$k$ dispersion becomes nearly flat, indicating a heavy composite object.
The peak position at $k\simeq 0$, $\omega_{\rm peak}$, provides a direct estimate of the quartet binding energy through $\omega_{\rm peak}-3\mu\simeq 0.33$, which is in the order of $E_{\rm b}=1/a^2\equiv 0.25$.
This behavior is characteristic of the strong-coupling side of the crossover, where fermionic excitations are strongly damped and the excitation scale is governed by the bound-state formation rather than by coherent Bogoliubov quasiparticles in the BCS pairing theory.

The spectral evolution shown in Fig.~\ref{Fig:2} is consistent with earlier Green's-function studies of quartet condensation in nuclear matter~\cite{Sogo2010Phys.Rev.C81.064310,schuck2012quartetting}, in which the analytical expression of $G_k(\omega)$ was reported.  
Those studies showed that the quartet contribution couples a single-particle excitation via a three-hole continuum. 
The integration of the relative momenta of the three holes generates a sizable imaginary part of the self-energy and broadens the single-particle spectrum as we showed in Fig.~\ref{Fig:2}.  
In this way, the quasiparticle pole loses spectral weight and becomes broader when the interaction strength increases. 
The main signature of the spectrum in the crossover regime is the transfer of weight from the quasiparticle pole to the incoherent continuum without the disappearance of the quartet order parameter. 
Our results thus extend the earlier nuclear-matter analyses by resolving $A(k,\omega)$ throughout the one-dimensional crossover.

\begin{figure}
  \includegraphics[width=0.45\textwidth]{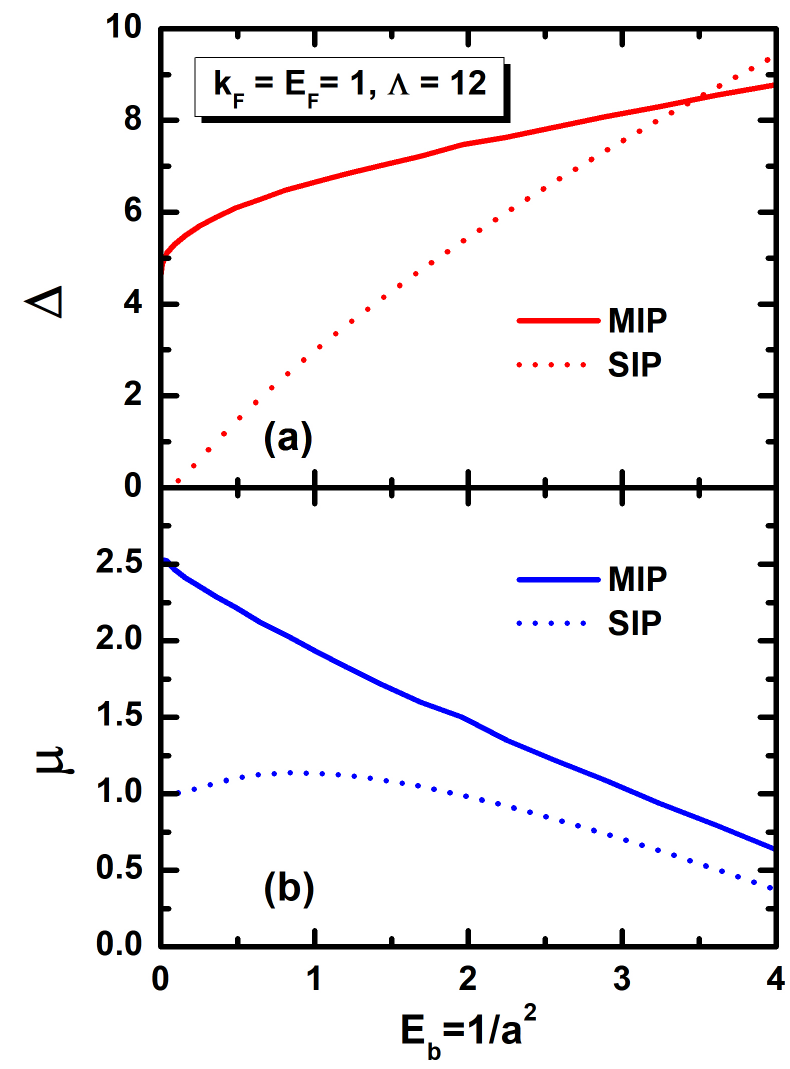}
  \caption{
  (a)~Quartet order parameter $\Delta$,
  and (b)~chemical potential $\mu$ as functions of the four-body binding energy $E_{\rm b}$ obtained within the SIP and MIP ansatzes.
  In the entire regime,
  the number density is fixed at $4/\pi$ with $k_{\rm F}=E_{\rm F}=1$.
  }\label{Fig:SIP}
\end{figure}

Finally, we compare $\Delta$ and $\mu$ obtained from the SIP and MIP ansatzes at the fixed density $\rho=4/\pi$ in Fig~\ref{Fig:SIP}.
In the weak-binding limit $E_{\rm b}\to0$, the SIP ansatz yields a strongly suppressed $\Delta$ and a reduced shift of the chemical potential compared with the MIP result, reflecting its inability to capture the broad internal momentum correlations of extended quartets near the crossover regime.
In particular, at $E_{\rm b}\rightarrow 0$, the SIP ansatz leads to $\Delta\rightarrow 0$ and $\mu\rightarrow 1$, that is, a non-interacting limit, in contrast to the MIP result. 
This difference originates from the renormalization scheme where the squared bare coupling $g_0^2$ is proportional to $E_{\rm b}$ in the SIP ansatz. 
On the other hand, in the MIP ansatz, the limit of $E_{\rm b}\rightarrow 0$ corresponds to the unitarity limit where quantum fluctuations play a crucial role.
With increasing $E_{\rm b}$, these two results gradually approach each other as the physics becomes dominated by compact, tightly bound quartets for which the detailed internal momentum structure plays only a subleading role.
In the regime $E_{\rm b}\gtrsim4$, however, the SIP order parameter exceeds the MIP value, indicating an overconfinement of internal degrees of freedom due to the restricted $\pm k$ structure.
In contrast, the MIP ansatz retains the full internal momentum space of the quartet and therefore exhibits a more moderate growth of $\Delta$, providing a physically meaningful description of genuine four-body binding.
Incidentally, it should be noted that the region with larger $E_{\rm b}$ is quantitatively sensitive to the choice of $\Lambda$. Nevertheless, an important point of our comparison is that the difference between MIP and SIP ansatzes does not induce qualitative differences on $\Delta$ and $\mu$ in such a regime.

\section{Discussion and perspectives}\label{sec:VII}

In this work, we have presented a comparative study of quartet-superfluid-state description between the quartet BCS variational framework and the field theoretical approach based on the generalized Nambu–Gor'kov formalism. 
Taking a one-dimensional four-component Fermi gas into account as a controllable benchmark, we have discussed the microscopic requirements for a consistent description of quarteting in fermionic matter.

It is seen that when the full internal momentum structure of the four-body bound state is retained, the quartet BCS theory reproduces the exact four-body result in the dilute limit and yields a quartet gap equation consistent with that derived from the generalized Nambu–Gor'kov formalism. 
This establishes a direct conceptual connection between variational quartet condensation pictures and field-theoretical descriptions.

As for the single-particle excitation spectrum, while no hard excitation gap emerges across the crossover from a quartet superfluid to a tetramer-dominated regime, fermionic quasiparticles are progressively suppressed by strong multiparticle damping associated with three-hole intermediate states, resulting in a continuum-dominated spectral response even in the presence of a finite quartet order parameter. 
The crossover is therefore governed by the loss of quasiparticle coherence and the emergence of few-body bound-state physics.

Finally, we have compared the numerical results of the quartet order parameter and the chemical potential obtained from the different ansatzes (i.e., SIP and MIP) in the quartet BCS theory.
It has been confirmed that their behaviors are qualitatively insensitive to the choice of SIP and MIP ansatzes in the strong-coupling region.
On the other hand, there is a deviation between two ansatzes in the weak-coupling regime, reflecting the different renormalization schemes of the coupling constant.

The restriction to a pure four-body interaction in Eq.~\eqref{hamiltonian} provides a controlled benchmark for genuine quartet correlations. 
On the other hand,  pairwise interactions are also important in realistic systems. 
If an SU(4)-symmetric attractive two-body interaction is considered, the pair and quartet anomalous averages should in general be determined on an equal footing. 
At the few-body level, the two-body coupling introduces an additional physical scale. 
The binding energy is therefore no longer fixed solely by the four-body scattering length as in Eq.~\eqref{eq:7}, and dimer thresholds can affect the tetramer pole. 
In the variational formulation, antisymmetric pair amplitudes can be added to the quartet coherent state, as considered in Ref.~\cite{Guo2022Phys.Rev.C105.024317}. 
In the Green's-function formulation, the two-body interaction generates a Hartree contribution and the conventional pairing self-energy $\Sigma_{2}(k,\omega)\sim|\Delta_{2}|^{2}/(\omega+\xi_{-k})$. 
These terms must be treated together with the three-hole quartet self-energy $\Sigma_{4}(k,\omega)$, resulting in coupled self-consistent equations for the two order parameters. 
Depending on the interaction strengths and density, the pair and quartet condensates may compete or coexist. 
The two channels also affect the single-particle spectrum differently. 
At the mean-field level, the pair channel produces sharper and more particle--hole-symmetric Bogoliubov branches, whereas the quartet channel produces an asymmetric continuum. 
The pair channel may therefore partially restore coherent spectral weight relative to the pure quartet case. 
The same coupled framework can be formulated for the attractive SU(4) Hubbard model studied in Ref.~\cite{Wan2026arXiv2604.14289}. 
For a direct lattice treatment, the continuum dispersion is replaced by the lattice dispersion, while both pair and quartet order parameters are retained.

The present framework provides a basis for systematic extensions toward realistic interactions, higher densities, and applications to nuclear matter, neutron-rich nuclei, and multi-component ultracold atomic gases in higher dimensions. 
The field-theoretical description enables us to address the finite-temperature phase transition of the quartet superfluid state.
Furthermore, the reduction of numerical cost of applying quartet BCS theory with keeping important physical properties is also practically worthwhile to investigate.

\begin{acknowledgments}
This work benefited greatly from discussions with the late Peter Schuck, whose insights have had a lasting influence on this study.
The authors thank Youngman Kim, Masaaki Kimura, Tomoya Naito, and Sibo Wang for useful discussions.
This works is supported in part by the National Research Foundation of Korea (NRF), funded by Ministry of Science and ICT (RS-2024-00436392).
Y.G. is supported by RIKEN Special Postdoctoral Researchers Program.
H.T. acknowledges the JSPS Grants-in-Aid for Scientific Research under Grants No.~JP22H01158 and No.~JP22K13981.
H.L. acknowledges the JSPS Grant-in-Aid for Scientific Research under Grants No.~JP26K07063 and No.~JP26K01431.

\end{acknowledgments}

%

\end{CJK}
\end{document}